\documentclass[aps,pra,twocolumn]{revtex4-1}
\usepackage[usenames]{color}
\newcommand{\comment}[1]{}
\usepackage{graphicx}
\usepackage{epsfig}
\usepackage{amsmath}

\newcommand{\expect}[1]{\langle {#1} \rangle}

\def \be{\begin{equation}}
\def \ee{\end{equation}}
\def \ba{\begin{array}}
\def \ea{\end{array}}
\def \beq{\begin{eqnarray}}
\def \eeq{\end{eqnarray}}
\def \bed{\begin{displaymath}}
\def \eed{\end{displaymath}}

\usepackage{color}

\begin{document}
\title{Prethermalization in quenched spinor condensates}
\date{\today}
\author{Ryan Barnett,$^1$  Anatoli Polkovnikov,$^2$  and Mukund
  Vengalattore$^3$}

\affiliation{
$^1$Joint Quantum Institute and Condensed Matter Theory Center,
 Department of Physics, University of Maryland, College Park, MD
 20742, USA}
\affiliation{
$^2$Department of Physics, Boston University, 590 Commonwealth Ave.,
Boston, MA 02215, USA}
\affiliation{$^3$Laboratory of Atomic and Solid State Physics, Cornell University, Ithaca, NY 14853, USA}

\begin{abstract}
Motivated by recent experiments, we consider the dynamics of spin-one
spinor condensates after a quantum quench from the polar to
ferromagnetic state from varying the quadratic Zeeman field $q$.  We
apply the Truncated Wigner Approximation (TWA) to the spinor system, including
all spatial and spin degrees of freedom. For short times, we find
full agreement with the linearized Bogoliubov analysis. For longer
times, where the Bogoliubov theory fails, we find that the system
reaches a quasi-steady prethermalized state.  We compute the Bogoliubov spectrum
about the ferromagnetic state with general $q$ and show that the
resulting finite temperature correlation functions grossly disagree
with the full TWA results, thus indicating that the system does not
thermalize even over very long time scales.  Finally we show that the
absence of thermalization over realistic time scales is consistent
with calculations of Landau damping rates of excitations in the
finite-temperature condensate.
\end{abstract}
\maketitle

\section{Introduction}

Advances in the field of ultracold atomic gases have spurred great
interest in the non-equilibrium dynamics of quantum many-body
systems. The ability to engineer paradigmatic model Hamiltonians, the
near perfect isolation from the environment and the experimentally accessible
timescales of evolution have made it possible to address fundamental
questions about the dynamics of closed, interacting quantum many-body
systems.

Of particular interest in this context is the study of a `quench'
across a quantum phase transition. Here, one or more parameters of the
Hamiltonian are changed rapidly resulting in a non-equilibrium
evolution of the quantum system towards the establishment of long
range order. This evolution is accompanied by a spatially
inhomogeneous symmetry breaking and the formation of topological
defects seeded by the quench. Accurate, time-resolved studies of such
quenches are of fundamental importance to a wide range of
non-equilibrium phase transitions.

Recently, an instance of such a quench across a phase transition has
been experimentally realized with quantum degenerate spin-1 Bose gases
of $^{87}$Rb \cite{sadler06,vengalatorre08}.  At low magnetic fields,
these multicomponent fluids are characterized by a contact interaction
that favors a ferromagnetic phase. At large external magnetic fields,
the quadratic Zeeman energy (QZE) dominates the interaction and favors
a paramagnetic (`polar') phase. As shown in Fig(1), these two phases
are separated by a continuous phase transition. 
In the experiment, {\em in situ} images of the spin textures following a
quench of the degenerate gas into the ferromagnetic phase revealed the
inhomogeneous growth of transversely magnetized domains accompanied by
the sporadic observation of topological defects that were characterized
as spin vortices.

Motivated by this experiment, here we consider the evolution of spin textures
following a quench from the polar phase to the ferromagnetic phase.
While the short-time growth of magnetization during such a quench has been
analyzed~\cite{lamacraft07, saito07, damski07,
  uhlmann07, cherng08, baraban08, klempt09, saito07b, sau09}, here we consider the evolution
  over much longer periods and study the manner in which the spin degrees of
  freedom thermalize following the quench. Applying the truncated Wigner
  approximation (TWA) to this problem (for an overview of the TWA method and
  its applications, see Refs.~\cite{blakie08, polkovnikov10}), we analyze the
  long time dynamics of the spin degrees of freedom.

We find that unless the QZE is quenched to the immediate vicinity of the critical
value corresponding to the transition to the ferromagnetic phase, the system
reaches a slowly evolving quasi-steady state with exponentially decaying
correlations. This correlation length very slowly increases
in time in a manner consistent with coarsening dynamics
\cite{bray94}. 
We also find and interpret previously unexplored physics of the dynamics of spinor
condensates such as the emergence of longitudinal magnetization.

\section{The Hamiltonian and Mean-Field Phases}

Following the experimental situation, we consider the parameter regime characteristic
of $F=1$ spinor condensates of $^{87}$Rb. Also, like in the experiment, the gases are
confined in a quasi-2D geometry wherein the spatial extent $d_y$ of the condensate along
one dimension is less than the spin healing length. This condition implies that the
spin dynamics along this axis are effectively frozen. The starting point is the
Hamiltonian
\begin{equation}
\mathcal{H} = \mathcal{H}_0  + \mathcal{H}_{\rm int}
\label{Eq:H}
\end{equation}
where the free Hamiltonian is
\begin{equation}
{\cal H}_0 = \int d^2r \Psi^\dagger \left(-\frac{\hbar^2}{2m}\nabla^2
+ V + q f_z^2 \right) \Psi
\end{equation}
and $\Psi=(\psi_1,\psi_0,\psi_{-1})^T$ is a three component spinor,
$V$ is the trapping potential, $f_{x,y,z}$ are the spin-one matrices,
and $q$ is the quadratic Zeeman shift.  The
interaction Hamiltonian is
\begin{equation}
{\cal H}_{\rm int} = \int d^2r \left( \frac{1}{2} c_0 n^2 + \frac{1}{2}
c_2 F^2 \right).
\end{equation}
Here we have $n=\Psi^\dagger \Psi$, ${\bf F} = \Psi^\dagger {\bf f}
\Psi$, and the parameters $c_0$ and $c_2$ can be expressed in terms of
the $s$-wave scattering lengths as
$c_0= \frac{4\pi \hbar^2}{d_y 3 m}(a_0+2a_2)$ and $c_2=\frac{4\pi
  \hbar^2}{d_y 3m}(a_2-a_0)$.  There is also a linear Zeeman shift in the system proportional to
$F_z$ but this can be dropped since $F_z$ commutes with the
Hamiltonian and thus is conserved. Spinor gases of $^{87}$Rb are
characterized by the scattering lengths $a_0 (a_2) = 101.8 (100.4)$ Bohr radii
respectively. Because $a_2 < a_0$, the spin dependent interaction in these
spinor gases favors the ferromagnetic state. Since for these  parameters $c_0 \gg |c_2|$, fluctuations in the total density are suppressed. Further, due to the weak spin interaction strength, the interparticle separation at typical densities is far smaller than the spin healing length, thereby suppressing the role of
quantum depletion.

\begin{figure}
\includegraphics[width=3in]{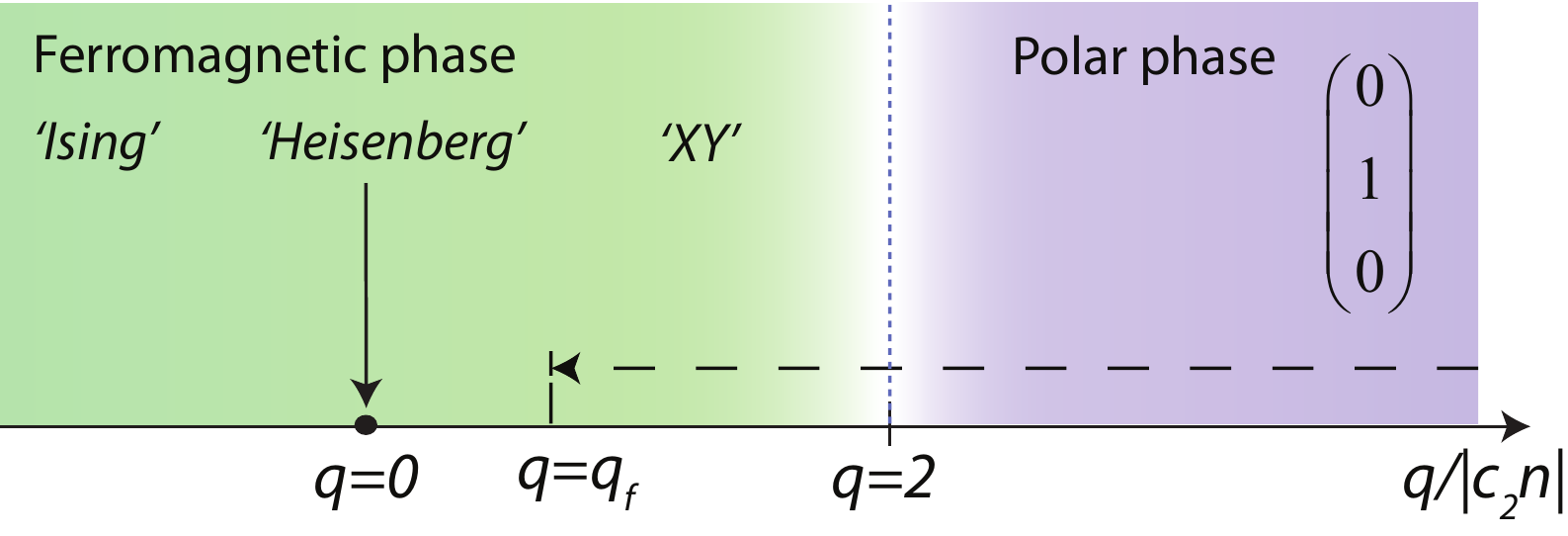}
\caption{(Color Online) Zero-temperature phase diagram of a spin-1 Bose condensate
indicating the transition from a polar phase to a ferromagnetic phase
at low quadratic Zeeman energy (QZE). The dashed line indicates the direction
of the quench from an initial QZE $q \gg 2 |c_2| n_0$ to a final QZE $q=q_f$.}
\label{Fig:phasediagram}
\end{figure}

\section{Short-Time Theory}

As illustrated in Fig.~\ref{Fig:phasediagram}, for $q>q_0\equiv 2|c_2| n_0$, the ground
state is the polar state while decreasing $q$ below $q_0$, the ground state
acquires a magnetic moment and eventually reaches the fully polarized
ferromagnetic state at $q=0$.  Initially, we take $q=q_i$ to be the
largest energy parameter in the Hamiltonian and quench to a final
state $q=q_f$ where $0\le q_f<q_0$.  For short times, the dynamics is expected
to be well-described by expanding the Hamiltonian to quadratic order
around the initial polar state \cite{lamacraft07, saito07, damski07,
  uhlmann07, cherng08, baraban08, klempt09},
$\Psi=\sqrt{n_0}(0,1,0)^T$. 
Under this expansion the Hamiltonian becomes
\begin{align}
{\cal H} &\approx \sum_{\bf k} \left( (\varepsilon_{\bf k}+ c_2 n_0 +q_f) (
\psi_{1,{\bf k}}^\dagger\psi_{1,{\bf k}}+
\psi_{-1,{\bf k}}^\dagger\psi_{-1,{\bf k}})
\right. \notag
\\  & \qquad \left.
+
c_2 n_0 ( \psi_{1,{\bf k}}^\dagger \psi_{-1,-{\bf k}}^\dagger+
\psi_{-1,-{\bf k}} \psi_{1,{\bf k}})
\right)
\end{align}
where we have dropped  the term describing the stiff density fluctuations. For
simplicity,  we have considered the system in the continuum, in
the absence of the trapping potential. In the ferromagnetic regime this quadratic Hamiltonian gives modes with imaginary frequencies, which correspond to unstable, exponentially growing in time modes~\cite{lamacraft07}.

To quantify the spin dynamics after the quench, it is natural to consider
transverse and longitudinal magnetization correlation functions
\begin{align}
\label{Eq:cf}
G_{\perp}({\bf r}, t) &= \frac{1}{n_0^2} \expect{:F_{+}({\bf r} )F_{-}(0):} \\
G_{z}({\bf r}, t) &= \frac{1}{n_0^2} \expect{:F_z({\bf r}) F_z(0):}
\end{align}
\cite{lamacraft07} and the concomitant gain functions, $G_{\perp}(t)$ and $G_{z}(t)$,
which are the above evaluated for ${\bf r}=0$.  For $q_0t /\hbar\gg 1$
and $q_f = 0$, the above correlations within the linearized theory give
\begin{align}
G_{\perp}(t)&=\frac{1}{\sqrt{8\pi q_0 t/\hbar}} \frac{1}{n_0 \xi_s^2}e^{q_0 t/\hbar} \\
G_{z}(t)&=\frac{1}{64\pi q_0^2 t^2\hbar^2} \frac{1}{(n_0 \xi_s^2)^2} e^{2q_0 t/\hbar}
\end{align}
where $\xi_s=\sqrt{\hbar^2/q_0 m}$ is the spin coherence length.
Interestingly, the longitudinal gain grows with twice the exponent
of the transverse gain. However, the longitudinal magnetization is suppressed by a large
factor $1/(n_0 \xi_s^2) \sim 10^{-4}$ for the experimental parameters of Ref.~\cite{sadler06}.
Thus, $G_z(t)$ remains much smaller than $G_\perp(t) $ up to the times
where the latter saturates 
$G_{\perp}(t)\approx 1$ and the linearized theory does not work (see Fig.~\ref{Fig:mag_growth}).

\section{Truncated Wigner Simulations}

As seen from the exponential growth of the gain functions, the
above theory clearly fails once the transverse magnetization is of
order unity.  To gain a more complete understanding we turn to
truncated Wigner simulations.  This technique involves propagating
the full spinor Gross-Pitaevskii Equations (GPEs) of motion obtained by
taking the classical limit of Eq.~(\ref{Eq:H}) seeded with random initial conditions for canonical degrees of freedom distributed according to the Wigner transform of the initial state. For our case the initial state is taken to be the
vacuum of Bogoliubov quasiparticles for a theory expanded about the
polar state.  For the limiting case of $q_i = \infty$ this
has the particularly simple form:
$
W({\Psi})\approx \delta(|\psi_{0,0}|^2-n_0)\prod_{\bf k}
8\exp\left[-2\sum_{\alpha}|\psi_{\alpha,{\bf k}}|^2\right],
$
where the factor of $8$ ensures the correct normalization of the
Wigner function  \cite{polkovnikov10}.

To propagate the wave functions governed by the spinor GPE, we first
discretize the system effectively by using a lattice with the constant
$\ell$ chosen such that $ \ell \ll \xi_s$ and $n_0 \ell^2\gg 1$ and then use a
split-operator method that is accurate up to cubic order in the time
step size \cite{javanainen06,barnett10}. The TWA is an approximate method
resulting in the leading order in the expansion of classical dynamics
in quantum fluctuations~\cite{polkovnikov10}. It is also known to be
asymptotically accurate at short times and exact for quadratic
theories. In our case we expect this method to be quantitatively
accurate at all times. Indeed the small parameter of the expansion is
$1/(n_0\xi_s^2)\ll 1$ so that quantum corrections to TWA are suppressed
by this factor and are potentially only important at very long
times. At longer times when nonlinearities become relevant, the
momentum modes become highly occupied and the dynamics remains
classical. Putting the system on a lattice ensures that there is no
potential spurious effect from vacuum occupation of high-energy modes,
which sometimes impedes the validity of TWA at long
times~\cite{blakie08}. We checked that the results of our simulations
are insensitive to the choice of the cutoff as long as it is shorter
than the spin coherence length.

\begin{figure}
\includegraphics[width=3in]{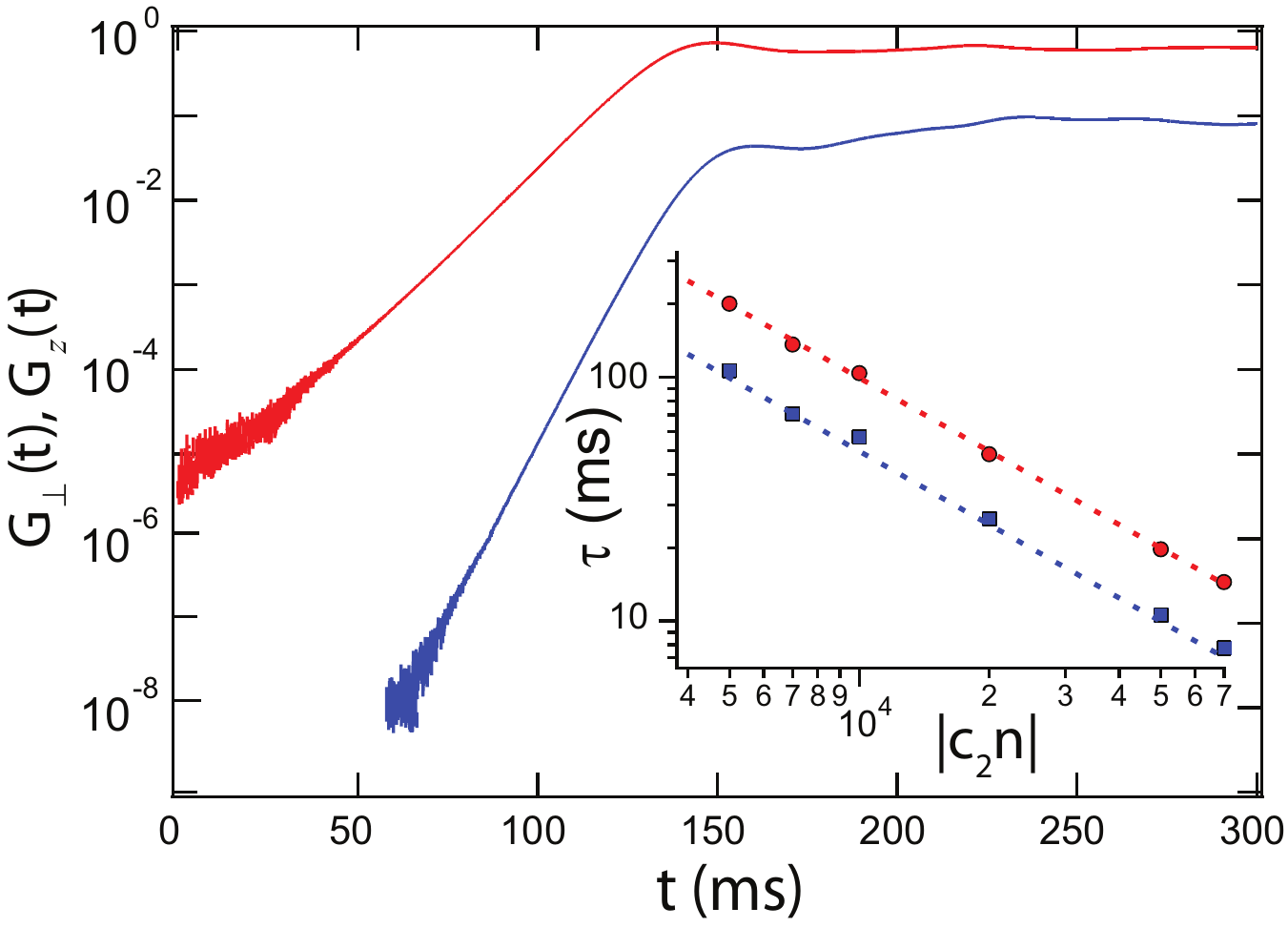}
\caption{(Color Online) The growth of transverse (red) and longitudinal (blue) magnetization following
a quench to $q_f=0$. Inset: The characteristic time constants for the growth of
transverse and longitudinal magnetization densities versus the 2D interaction strength
$|c_2 n|$. The dashed lines indicate the predictions of the linearized
Bogoliubov treatment for the respective parameters.}
\label{Fig:mag_growth}
\end{figure}

Shown in Fig.~\ref{Fig:mag_growth} are the TWA results for the
transverse and longitudinal gain functions in the absence of a trap
compared with the analytic results based upon the above linearized
theory.  
At short evolution times following the quench, the
simulations predict the exponential growth of both the transverse and
longitudinal magnetization densities in a manner that is in excellent
agreement with the linearized theory. During this period, the spin
textures are characterized by ferromagnetic domains that are predominantly
oriented in the transverse plane.  

\section{Long-time behavior}

We now move on to discuss the long-time dynamics of the quenched
spinor condensate.  As is seen in the gain functions in
Fig.~\ref{Fig:mag_growth}, the magnetization of the system eventually
saturates and reaches a steady state.  This motivates one
to consider the prospect of thermalization in the system.  More
specifically, under the assumption of thermalization, the system will
evolve to the ground state at $q=q_f$, with the excess energy
accounted for by a superposition of elementary excitations about this
configuration.  We therefore will consider the theory about the state
at $q=q_f$.

We define the heating (excess energy) of the system as
\begin{equation}
Q=\expect{\cal H}_i - \expect{\cal H}_{\rm gs}
\end{equation}
where the above expectation values are evaluated for the initial state
immediately after the quench and the ground state for $q=q_f$. To
extract the equilibrium temperature this heating can be compared with
the thermal energy of the system. It is convenient to use the
following parameterization of the spinor
\begin{equation}
\label{Eq:Psi}
\Psi = \sqrt{n} e^{i\alpha}\left( \sin\eta \cos\phi e^{i\xi},
  \cos\eta e^{i\gamma}, \sin \eta \sin \phi e^{-i\xi} \right)^T.
\end{equation}
With this, one can see that for $0<q_f/q_0< 1$ the classical energy is
minimized for $\phi=\pi/4$, $\gamma=0$, and $\cos(2\eta) = q_f/q_0$.
One then obtains that the above heating is
\begin{equation}
Q=\frac{1}{4}N q_0\left(1-q_f/q_0\right)^2
\label{q}
\end{equation}
where the details of this calculation are in Appendix \ref{A1}.  Note
that the above is based on classical energy differences. In addition
to this classical contribution to the heating there is also a
quantum correction coming from the zero point fluctuations of the
condensate. However this contribution is suppressed by a large factor
$\sqrt{n_0\xi_s^2}\sim 100$ and thus is not important.

The thermal energy of the final state is found from
\begin{equation}
Q=U(T) \equiv {\bf Tr} \left( e^{-{\cal H}/T}{\cal H} \right)/Z
\end{equation}
where $Z={\bf Tr}\left( e^{-{\cal H}/T}\right)$ is the partition
function and the ground state energy of the system is set to be zero.
To evaluate the above we develop a second Bogoliubov theory by
expanding the Hamiltonian to quadratic order about the minimum at
$q=q_f$. Note that within the chosen parametrization one automatically
avoids subtleties related to the lack of a long range order in two
dimensions at finite temperatures. The quadratic
theory is then self-consistently checked by verifying that the
resulting thermal depletion is small for the computed temperature.

The full Bogoliubov analysis of the spectrum linearized around the
ferromagnetic minimum is rather cumbersome because of the
phonon-magnon coupling~\cite{uchino10} (which vanishes for the special
points $q_f=0$ and $q_f=q_0$). However, $^{87}$Rb has a natural
separation of energy scales since $c_0\gg |c_2|$ which inhibits density
fluctuations.  This significantly simplifies the Bogoliubov analysis
and effectively results in the decoupling of the stiff density fluctuation
and the spin modes. In this limit
one finds the simplified spectrum consisting of two modes with the
dispersion (see Appendix \ref{A1})
  \begin{align}
\label{Eq:gapless}
\omega^{(\perp)}_{\bf k} &= \sqrt{\varepsilon_{\bf k}  (\varepsilon_{\bf k}  + q_f)}\\
\omega^{(z)}_{\bf k}  &= \sqrt{(\varepsilon_{\bf k}  + q_0) \left(\varepsilon_{\bf k}  +
    \frac{q_0^2-q_f^2}{q_0}\right)} \notag
\end{align}
where $\varepsilon_{\bf k}$ is the free particle dispersion.  Note that
these reduce to the known dispersions in the limits of $q_f = 0$ and
$q_f = q_0$ \cite{ho98,ohmi98}.  Furthermore, the large $c_0$ limit of
the full dispersion \cite{uchino10} agrees with Eq.~(\ref{Eq:gapless}).
These expressions are used to compute the thermal energy of the
system. It is straightforward to verify that unless $q_f$ is very
close to $q_0$,  $1-q_f/q_0 \not\gg 1/\sqrt{n_0\xi_s^2}$, the dominant
contribution to the thermal energy $U(T)$ comes from the quadratic
dispersion of these modes: $\omega^{(\perp)}_{\bf k}\approx
\omega^{(z)}_{\bf k}\approx \varepsilon_{\bf k}$. This can be justified \emph{a
  posteriori}  by comparing the temperature with $q_0$. With these conditions,
$
U(T)\approx N\pi T^2/ 6 (n_0 \xi_s^2) q_0 .
$
Finally, equating this to the heating, one finds
\begin{equation}
\label{Eq:T}
T= \sqrt{\frac{3}{2\pi} n_0 \xi_s^2} (q_0-q_f).
\end{equation}

To test the hypothesis of thermalization, we use the developed
quadratic theory  to evaluate finite temperature correlation functions
(evaluated at temperature Eq.~(\ref{Eq:T})) and compare with long-time TWA
results.  The most natural correlation function to consider is
$G_{\perp}$ defined in Eq. (\ref{Eq:cf}).  The gapless mode, Eq.~(\ref{Eq:gapless}), which
leads to algebraic decay of the correlation function in two dimensions
has the dominant contribution.  Within the quadratic Bogoliubov theory, the
correlation function is found to be (see Appendix \ref{A2})
\begin{equation}
\label{Eq:cflong}
G_{\perp}({\bf r}) = \left[1-\left(q_f/q_0\right)^2\right]   (r/\xi_s)^{-\alpha}
\end{equation}
with exponent $\alpha=\sqrt{6/\pi^3} \cdot \sqrt{1/ n_0 \xi_s^2}$.
Because of the dependence of the exponent $\alpha$ on the small
parameter $1/n_0 \xi_s^2$, we see that, assuming equilibrium,
$G_{\perp}({\bf r})$ will not decay by any appreciable amount over the relevant length scales of
the condensate, which are typically of the order of tens or hundreds
of the spin coherence length $\xi_s$. For the same reason in
equilibrium the system should not have any vortices.

We  point out that the resulting thermal state of the condensate
with temperature $T$ given by Eq.~(\ref{Eq:T}) has some interesting
properties.  On the one hand except for a small vicinity of
$q_f\approx q_0$ it is a high-temperature classical state
characterized by a temperature much higher than the chemical
potential. This implies that quantum depletion in this state is
negligible. On the other hand, the smallness of the exponent $\alpha$
shows that this state is well below the Kosterlitz-Thouless (KT) transition
temperature due to the unbinding of vortices in the spin degrees of
freedom.   
More specifically, the KT transition temperature scales as $T_{KT}
\sim q_0 n_0 \xi_s^2$ which can be compared with Eq.~(\ref{Eq:T}).
In this respect, this is a low-temperature state.

\begin{figure}
\includegraphics[width=3in]{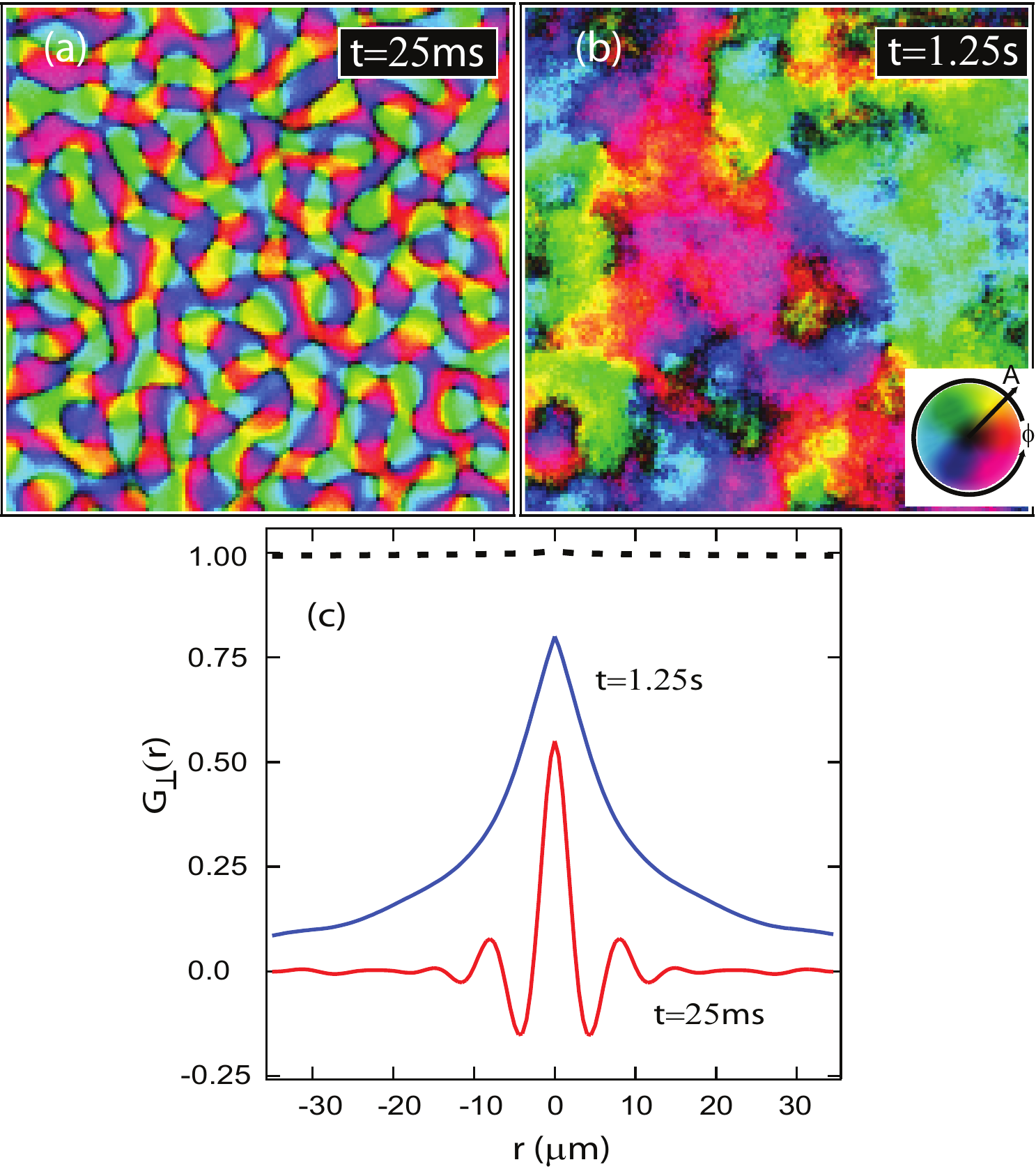}
\caption{(Color Online) The transverse magnetization densities for short (a) and long
  (b) times for $q_f = 0$ over a $70\times70 \mu{\rm m}$ region. 
The amplitude and orientation of the
  transverse magnetization are indicated by the brightness and hue
  (color wheel shown).  (c) The spin correlation function for these
  same times.  While the correlation function at short times is in
  excellent agreement 
with Bogoliubov theory, the decay of this function at long evolution
times is much more rapid than predicted by 
Eq.~(\ref{Eq:cflong}) (dashed line).}
\label{Fig:thermal}
\end{figure}

We now compare these predictions to the results of the TWA simulations
for long times.  Shown in Fig.~\ref{Fig:thermal} are magnetizations
for short and long times as well as correlation functions after the
quench.  Displayed are results for zero final quadratic Zeeman field
$q_f=0$, though qualitatively very similar results occur for finite
values of $q_f$ (for a full animation of the dynamics see EPAPS
material \cite{barnett11}).  As shown in Fig.~\ref{Fig:thermal}(c), the
correlation function $G_{\perp}$ before saturation for short times has
the functional dependence of a Bessel function $J_0(r/\xi_s)$ as
previously predicted \cite{lamacraft07}.  For long times, the
correlation function reaches a steady state and decays by over a
factor of five from its $r=0$ value for a system size on the order of
$50$ spin coherence lengths.  Such decay is qualitatively incompatible
with the theory developed by assuming thermalization.  We thus
conclude that these spinor condensate systems do not thermalize at
appreciable time scales but rather reach a quasi-steady
(prethermalized) regime which evolves anomalously slow in time. Such
prethermalized phases were suggested earlier in the contexts of weakly
interacting fermions~\cite{moeckel_10, eckstein_kollar_09}, Bose
superfluids~\cite{mathey_10,kitagawa11}, and dynamics of the early
Universe~\cite{berges_borsanyi_04}. What is perhaps unexpected is that
in this system anomalously slow relaxation occurs at very high levels
of the heating in the system.

To support these results, it is useful to compare with the Landau
damping rates of long-wavelength Bogoliubov modes as a result of
scattering from thermal modes
\cite{lifshitz81,pitaevskii97,giorgini98,pethick08}.  Such an analysis
is carried out in Appendix \ref{A3}.  It is found that typical
lifetimes of these modes is on the order of $100$ seconds.  This rate
is considerably lower than those of scalar condensates due to the weak
spin-dependent interaction and small thermal depletion.  The lifetime
estimates provide lower bounds for the thermalization time and are
consistent with the numerical results.  However, such long time scales
are outside of our numerical as well as experimental access.

In conclusion, we have examined the long time evolution of quantum
degenerate spinor gases following a quench to a ferromagnetic
phase. Assuming a thermalized final state, we find that the finite
temperature spin correlations are characterized by an algebraic decay
over length scales much larger than the relevant length scales of the
condensate.  In distinct contrast, numerical simulations based on the
truncated Wigner approximation indicate a rapidly decaying correlation
function even at extremely long evolution times.  This inconsistency
leads one to the conclusion that the quenched spinor condensates will
not thermalize over experimentally relevant timescales.  These results
are consistent with the Landau damping rates.  The role of dipolar
interactions \cite{vengalatorre08, vengalattore10} in the long time
evolution has yet to be examined.

\acknowledgements
For valuable discussions, we thank E. Altman, A. Lamacraft, L. Mathey, G. Refael
J. D. Sau, and A. Turner.
M. V. thanks L. M. Aycock and S. Chakram for valuable discussions and critical comments
on the manuscript.  We gratefully acknowledge financial support from
the NSF JQI Physics Frontier
Center and the Sherman Fairchild Foundation (R.B.), AFOSR
FA9550-10-1-0110 and the
Sloan Foundation  (A.P.), and Cornell University (M.V.). The authors
also thank the Aspen Center for Physics  for hospitality, where a part of this work was completed.

\appendix

\section{Bogoliubov analysis in the regime $0<q_f<2|c_2| n_0$}
\label{A1}
The easiest way to perform the Bogoliubov analysis of spinor
condensates is to use a parametrization of the spinors though the
density-angle variables as in Eq. (\ref{Eq:Psi}).  In this case one
avoids issues related to absence of the true condensation in
one-dimension at finite temperatures. Also the normal modes have a
very transparent physical meaning. The analysis of this Appendix very
closely mimics an analysis for the description of the low-energy
excitations of bosons in an optical lattice close to the
superfluid-insulator transition in the effective spin-one
representation~\cite{ehud_thesis, altman_02}.  In this
parametrization, the magnetization $F_z=\Psi^\star f_z\Psi$ and
$F_{z,2}=\Psi^\star f_z^2\Psi$ are given by
\beq
F_z&=&|\Psi_1|^2-|\Psi_{-1}|^2=n\sin^2(\eta)\;\cos (2\phi),\\
F_{z,2}&=& |\Psi_1|^2+|\Psi_{-1}|^2=n\sin^2(\eta).
\eeq
Similarly the square of the transverse magnetization $|F_\perp|^2=F_y^2+F_x^2$ reads:
\be
|F_\perp|^2={n^2\over 2}\sin^2 (2\eta) [1+\sin(2\phi)\cos(2\gamma)].
\ee
With these expressions, one finds for the interaction energy density (here
we include the quadratic Zeeman term in the interaction energy)
\beq
\mathcal H_{\rm int}&=&{c_0\over 2} n^2+ {c_2\over 2}(F_z^2+|F_\perp|^2)+q F_{z,2}\nonumber\\
&=&{c_0\over 2}n^2+c_2n^2\sin^2(\eta)\cos^2(\eta)[1+\sin(2\phi)\cos(2\gamma)]\nonumber\\
&+&qn\sin^2(\eta)+{c_2 n^2\over 2}\sin^4(\eta)\cos^2(2\phi).
\label{h_int}
\eeq
It is straightforward to check that for $c_2<0$ and $0<q<2|c_2|n=q_0$
the energy is minimized when $\phi=\pi/4$, $\gamma=0$, and
$\eta=\bar{\eta}$ where we define $\bar{\eta}$ according to
$q=q_0\cos(2\bar{\eta})$. There is an equivalent minimum obtained by
gauge transformation where $\phi\to \phi+\pi/2$ and
$\gamma\to\gamma+\pi$. The minimum of the interaction energy density is then
\be
E(q) \approx - q_0 n\sin^4(\bar{\eta})=-{g_0 n\over 4} \left(1 -{q\over q_0} \right)^2.
\ee
Note that in this expression we ignored the zero point energy
associated with the depletion, which is suppressed by a large factor
$1/\sqrt{n_0\xi_s^2}$.

Within the Bogoliubov approximation we need to expand the energy
around the minimum. For this purpose we define small deviations of the
angles $\phi$ and $\eta$ from the optimal values
$\phi=\pi/4+\delta\phi$, $\eta=\bar{\eta}+\delta\eta$ and do a second
order expansion in $\delta\phi$, $\delta\eta$, $\gamma$, and
$\xi$. Since we are interested in the limit $c_0\gg c_2$ the density
fluctuations are suppressed and we can set $\delta n=0$ and fix the
density at its equilibrium value $n=n_0$.  Then the expression for the
interaction energy density~(\ref{h_int}) reduces to
\begin{eqnarray}
&&\delta\mathcal H_{\rm int} \approx q_0 n_0 \sin^2(\bar{\eta})\nonumber\\
&&\times\left[\cos(2\bar{\eta})(\delta\phi)^2+\cos^2(\bar{\eta}) \gamma^2 +4\cos^2(\bar{\eta}) (\delta\eta)^2\right].
\end{eqnarray}
Under the same approximation the kinetic energy density term reads
\beq
&&\mathcal H_{\rm kin}\approx {\hbar^2\over 2m}\biggl[(\nabla\delta \eta)^2+\sin^2(\bar{\eta}) (\nabla\phi)^2
+\sin^2(\bar{\eta}) (\nabla\xi)^2\nonumber\\
&&+\cos^2(\bar{\eta}) (\nabla\gamma)^2+(\nabla\alpha)^2+2\nabla\alpha\nabla\gamma\cos^2(\bar{\eta})\biggr].
\eeq
And finally to determine the canonically conjugate variables we need
to write the Berry phase term, $i\Psi^\dagger \dot\Psi$ in the
linearized approximation:
\be
2n_0 \sin^2(\bar{\eta})\delta \phi\,
\dot\xi+n_0\sin(2\bar{\eta})\delta\eta\,\dot\gamma-\delta n[\dot\alpha+
\cos^2(\bar{\eta})\dot\gamma].
\ee
In this form it is clear that the phase conjugate to the density
fluctuations is $\tilde\alpha=\alpha+\cos^2(\bar{\eta})\gamma$. Then
the kinetic energy density can be rewritten as
\beq
&&\mathcal H_{kin}\approx {\hbar^2\over 2m}n_0\biggl[(\nabla\eta)^2+\sin^2(\bar{\eta}) (\nabla\phi)^2
+\sin^2(\bar{\eta}) (\nabla\xi)^2\nonumber\\
&&+(\nabla\tilde\alpha)^2+\cos^2(\bar{\eta})\sin^2(\bar{\eta}) (\nabla\gamma)^2\biggr].
\eeq
One can further redefine variables 
\bed
\delta\phi\to\tilde\phi/(2\sqrt{n_0}\sin(\bar{\eta})),\quad\xi\to \tilde\xi/\sqrt{n_0}\sin(\bar{\eta}),
\eed 
\bed \gamma\to\tilde\gamma/\sqrt{n_0}\sin(2 \bar{\eta}), \quad \eta\to\tilde\eta/\sqrt{n_0}.
\eed
Then, the spin part of the Lagrangian density (Berry phase minus Hamiltonian) becomes
\beq
&&\mathcal L \approx \tilde\phi\, \dot{\tilde\xi}+\tilde\eta\dot{\tilde\gamma}-g_s \left[{\cos(\bar{\eta}/2)\over 4}\tilde\phi^2+\cos^2\left({\bar{\eta}}\right) \tilde\gamma^2+\sin^2(2\bar{\eta}) \tilde\eta^2\right]\nonumber\\ &&-{\hbar^2\over 2m}\biggl[(\nabla\tilde\eta)^2+{1\over 4}(\nabla\tilde\phi)^2+(\nabla\tilde\xi)^2+\cos^2\left({\bar{\eta}}\right)(\nabla\tilde\gamma)^2\biggr].
\eeq
This Lagrangian density immediately gives two sets of normal modes
corresponding to oscillations in the $\phi-\xi$ variables, which
represent uniform rotations around $z$-axis, and in the $\eta-\gamma$
variables, which represent oscillations in the magnitude of the
transverse magnetization. The corresponding frequencies in the
momentum space are
\beq
&&\omega^{(\perp)}_{\bf k}=\sqrt{\varepsilon_{\bf k}(\varepsilon_{\bf k}+q_0 \cos(2\bar{\eta}))},\\
&&\omega^{(z)}_{\bf k}=\sqrt{(g_0 +\varepsilon_{\bf k})(\varepsilon_{\bf k}+q_0  \sin^2(2\bar{\eta}))}
\eeq
where $\varepsilon_{\bf k}=\frac{\hbar^2 k^2}{2m}$ is the free particle dispersion.  We note
that these expressions follow from taking $c_0\to\infty$ limit of the
general dispersion relations obtained in Ref.~\cite{uchino10}. There
are, however, advantages in using the density-angle representation since
it does not rely on the assumption of the existence of the true long
range order.

\section{Finite temperature correlation functions within the Bogoliubov theory}
\label{A2}

The Bogoliubov theory described in the previous section makes it
possible to compute correlation functions in two-dimensional systems.  
In the following we will consider the transverse magnetization correlation function
\begin{align}
G_{\perp}({\bf r}, t) &= \frac{1}{n_0^2} \expect{:F_{+}({\bf r}
  )F_{-}(0):} .
\end{align} 
Written in terms of the variables introduced in the previous section we have
\begin{equation}
F_{+} = \frac{1}{\sqrt{2}} n e^{i\alpha} \sin(2\eta)e^{-i\xi} \left(
  \cos(\phi) e^{i\gamma} + \sin(\phi) e^{-i\gamma}\right).
\end{equation}
According to the Mermin-Wagner theorem, a gapless mode will lead to a
power law behavior of correlation functions in two-dimensions.  The
gapless mode for our case corresponds to the $\phi-\xi$ variables.  It
is straightforward to see that these will dominate the correlation
functions.  Taking into account these two fluctuating fields one finds
\begin{equation}
<F_{+}({\bf r}) F_{-} (0) > = \sin^2(2\bar{\eta}) e^{-<\Delta
  \phi^2>/2}e^{-<\Delta \xi^2>/2}.
\end{equation}
In this equation, $\expect{\Delta \xi ^2}\equiv \expect{(\xi({\bf r})- \xi(0))^2}$ with a similar expression
for $\expect{\Delta \phi^2}$.

Using the analysis from above, one finds
\begin{equation}
\expect{\Delta \xi^2} = \frac{1}{N \sin^2(\bar{\eta})}\sum_{{\bf
    k}}\frac{\omega_{\bf k}^{(\perp)}}{\varepsilon_{\bf k}} \left(
  f(\omega_{\bf k}^{(\perp)}) + 1/2 \right) (1 - J_0(kr)).
\end{equation}
and
\begin{equation}
\expect{\Delta \phi^2} = \frac{1}{N
    \sin^2(\bar{\eta})}\sum_{\bf k}
\frac{\varepsilon_{\bf k}} {\omega_{\bf k}^{(\perp)}}\left(
  f(\omega_{\bf k}^{(\perp)}) + 1/2 \right) (1 - J_0(kr))
\end{equation}
where $f$ is the Bose distribution function.
It can be seen that the dominant contribution comes from $\expect{\Delta \xi^2}$ due to the
long-wavelength divergence of the sum over ${\bf k}$ which is cut off at $\sim 1/r$.  The
above expression is therefore well-approximated by
\begin{equation}
\expect{\Delta \xi^2}= \frac{T}{\pi \sin^2(\bar{\eta})
    q_0 n_0 \xi_s^2}\log(r/\xi_s).
\end{equation}
Using the expression for the temperature derived in the manuscript,
one arrives at Eq.~(\ref{Eq:cflong}).

\section{Landau Damping Rate}
\label{A3}

We take a spinor condensate with a single Bogoliubov
excitation of wave vector ${\bf k}$  and evaluate its lifetime $\tau_k$  due to scattering off of
short-wavelength thermal modes.  For simplicity we will consider the
gapless spin mode $\omega_{\bf k}^{\perp}$ as given in Eq.~(\ref{Eq:gapless}).
Such a rate is given by the
well-known Landau damping formula \cite{lifshitz81} which has been
generalized to Bose-Einstein Condensates in \cite{pitaevskii97,giorgini98,pethick08}
\begin{equation}
\frac{1}{\tau_k}=\frac{\pi}{\hbar} \sum_{\bf k'} |M_{{\bf k k'}}|^2 (
f(\omega_{\bf k}^{(\perp)}) - f(\omega_{{\bf k}+{\bf k'}}^{(\perp)}) ) \delta(\omega_{\bf k}^{(\perp)}+ \omega_{\bf
  k'}^{(\perp)}
-\omega_{{\bf k}+{\bf k'} }^{(\perp)}).
\end{equation}
where $f$ is the Bose-Einstein distribution
function.  In this equation, the matrix 
element $M_{{\bf k k'}}$ is given by \cite{pitaevskii97,giorgini98,pethick08}
\begin{equation}
M_{{\bf k k'}}= \frac{q_f}{2\sqrt{N}} \sqrt{\frac{\hbar k}{2ms}} \left(
  \frac{\varepsilon_{\bf k'}}{\omega^{(\perp)}_{\bf
      k'}}+\frac{\omega^{(\perp)}_{\bf k'}}{\varepsilon_{\bf k'}+q_f/2} \right)
\end{equation}
where $s$ is the sound speed of the mode $\omega^{(\perp)}_{\bf k'}$.
With this expression, the two-dimensional ${\bf k'}$-summation can be performed.  In
the limit $T \gg q_f$, the result is 
\begin{equation}
\frac{1}{\tau_k}=\frac{a_s k}{\hbar} \frac{T}{d_y / \xi_s}
\frac{q_f}{q_0} 1.13
= \frac{k \xi_s}{\hbar}\frac{T}{8\pi n_0 \xi_s^2} \frac{q_f}{q_0} 1.13
\end{equation}
where $a_s=(a_2-a_0)/3$ is the spin-dependent scattering length.  Using
the experimental parameters, and the expression for the temperature
given in Eq.~(\ref{Eq:T})  and taking $q_f=q_0/2$ we find that
\begin{equation}
\tau_k \approx 90{\rm s}
\end{equation}
for a mode having wave vector $q=1/(2\xi_s)$.


%

\end{document}